\documentclass [prb,twocolumn, amsmath,amssymb]{revtex4}
\setcitestyle{numbers,square}
\usepackage{appendix}
\usepackage{array}
\usepackage{amsmath,amssymb}
\usepackage{tabularx}
\usepackage{graphicx}
\usepackage{bmpsize}
\usepackage{bm}
\usepackage{color}
\usepackage{amssymb}
\usepackage [version=3]{mhchem}
\usepackage{newtxmath}
\usepackage{hyperref}
\usepackage{url}
\usepackage{physics}
\usepackage{here}
\usepackage{comment}
\DeclareMathAlphabet{\mathpzc}{OT1}{pzc}{m}{it}

\renewcommand{\vec}{\vb*}

\newcommand{\joya} [1]{\textcolor{red}{ #1}}

\begin{document}
\title{
Shift current response in twisted double bilayer graphenes
}
\author{Takaaki V. Joya}
\affiliation{Department of Physics, Osaka University, Toyonaka, Osaka 560-0043, Japan}
\author{Takuto Kawakami}
\affiliation{Department of Physics, Osaka University, Toyonaka, Osaka 560-0043, Japan}
\author{Mikito Koshino}
\affiliation{Department of Physics, Osaka University, Toyonaka, Osaka 560-0043, Japan}
\date{\today}

\begin{abstract}
We calculate the shift current response in twisted double bilayer graphenes (TDBG) by applying the perturbative approach to the effective continuum Hamiltonian.
We have performed a systematic study of the shift current in AB-AB and AB-BA stacked TDBG, where we have investigated the dependence of the signal on the twist angle, the vertical bias voltage and the Fermi level.
The numerical analyses demonstrate that the signal is greatly enhanced as the twist angle is reduced.
Notably, we also found that there is a systematic sign reversal of the signal in the two stacking configurations below the charge neutrality point for large bias voltages.
We qualitatively explain the origin of this sign reversal by studying the shift current response in AB-stacked bilayer graphene.
\end{abstract}

\maketitle
\section{Introduction}\label{sec:intro}

The study of moir\'{e} systems has been one of the major focusses of modern condensed matter physics.
By stacking two sheets with a relative twist angle, the total system shows a moir\'{e} pattern originating from the lattice mismatch between the two layers.
This moir\'{e} pattern leads to a modification of the band structure of the system, where the most notable example is the flat band formation and the experimental observation of superconductivity in twisted bilayer graphene (TBG) at an angle known as the magic angle $\theta \sim 1.1^\circ$~\cite{BistritzerMacDonald2011,CaoJarilloHerrero2018,CaoKaxiras2018}.
Studies has expanded to other graphene-based moir\'{e} systems, such as twisted monolayer-bilayer~\cite{RademakerAbanin2020,ParkJung2020,HeZhangYankowitz2021,XuNovoselov2021,ChenYankowitz2021,LiGao2022,TongYin2022,LiMao2022,BoschiPezzini2024,ZhangZhou2024}, twisted trilayer~\cite{ZhuKaxiras2020,ParkJarilloHerrero2021,MaXie2021,HaoKim2021,PhongGuinea2021,KimNadjPerge2022,TurkelPasupathy2022}, and other twisted multilayer graphene~\cite{LiuDai2019,ParkJarilloHerrero2022,NguyenCharlier2022,ShinMin2023}.

Twisted double bilayer graphene (TDBG) is a moir\'{e} system where two sheets of AB-stacked bilayer graphene (BLG) are stacked with a relative twist angle, as shown in Fig.~\ref{fig:3-structure}.
A variety of studies on TDBG has been performed, such as the observation of strongly correlated phenomena~\cite{Koshino2019,ChebroluJung2019,BurgMacDonald2019,CrosseMoon2020,LiuKim2020,HaddadiOleg2020,LiuDai2020,CulchacMorell2020,ShenZhang2020,CaoJarilloHerrero2020,HeYankowitz2021,SzentpeteriMakk2021,WangTuctuc2022,Ma2022,RubioVerduPasupathy2022,TomicEnsslin2022}, owing to its highly tuneable band structure, where a band gap can be opened through the application of a vertical bias voltage.
Importantly, TDBG can be fabricated in two distinct stacking structures, known as AB-AB and AB-BA, respectively shown in Fig.~\ref{fig:3-structure} (a) and (b).
The AB-AB stacking is composed of two bilayer sheets stacked 
and twisted with the same orientation, while the AB-BA stacking introduces a $180^\circ$ offset on the second bilayer.
A schematic diagram of the respective variants are shown in Fig.~\ref{fig:3-structure} (a) and (b).
It is known that the two variants have completely different valley Chern numbers while hosting very similar band structures~\cite{Koshino2019,ChebroluJung2019,LiuDai2019}.

Simultaneously, the shift current response has been under intense investigation due to its potential to probe the topology, quantum geometry and the symmetry of a variety of materials~\cite{TanzhengRappe2016,OrensteinMooreMorimoto2021,Aftab2022,MaSong2023,AhnNagaosaVishwanath2022,Cook2017,NagaosaMorimoto2017}.
This is a second order nonlinear optical response where light is rectified into dc current in noncentrosymmetric materials~\cite{vonBaltzKraut1981,Sipe2000,ParkerMorimoto2019}.
It is characterised by a quantity known as the shift vector, which is expressed in terms of the difference between the Berry connection of the initial and final band of the optical excitation. The shift current response has been experimentally measured in various 2D materials~\cite{Aftab2022,IbanezAzpiroz2020,ChangLouie2021,AkamatsuIdeueYoshii2021,ZhangMachida2022,ChangNagashio2023,PostlewaiteFiete2024,KitayamaOgata2024,FeiRappe2020}, where the two main types which have been actively studied are the family of transition metal dichalcogenides~\cite{Jiang2021,YangIdeue2022,DongYoshiiIdeue2023,TanXu2017,BritnellNovoselov2013,SchanklerRappe2021,HabaraWakabayashi2023,Kaner2020,Sauer2023,RangelFregoso2017,PandayFregoso2019} and ferroelectric 2D materials, such as transition metal monochalcogenides~\cite{Qian2023,ChangLouie2021,ChangNagashio2023} and CuInP$_2$S$_6$~\cite{Li2021,ZhangMachida2022}. 

Nowadays, the scope of investigation of the shift current has expanded to moir\'{e} materials~\cite{Hu2023,LiuDai2020,Chaudhary2022,ChenChaudhary2024,Ma2022,Kim2022,ZhangIdeue2019} where there is no need to rely on the polarisation of the monolayer to observe a finite shift current; for example, two non-polar materials with a lattice mismatch can be stacked in order to create an interface with broken inversion symmetry.
The shift current response in TBG, being one of the simplest moir\'{e} system, has been studied theoretically~\cite{LiuDai2020,Chaudhary2022,KaplanHolder2022,PenarandaOchoa2024}.
However, they generally host the $C_{2z}$ rotational symmetry, with the exception for certain commensurate stacking configurations (\textit{e.g.} $D_3$ structure~\cite{ChangSipe2022}), which renders the in-plane shift current to vanish~\cite{PenarandaOchoa2024}. 
To obtain a finite response, such symmetry must be broken. Approaches include introducing an asymmetric potential between the A and B sites through alignment with hBN~\cite{LiuDai2020,Chaudhary2022,KaplanHolder2022} or taking advantage of the spontaneous symmetry breaking at low temperatures.
There have also been studies of the shift current in other graphitic systems~\cite{LiuDai2020,Ma2022,ChenChaudhary2024} and heterostrucutres~\cite{BritnellNovoselov2013,Hu2023,AkamatsuIdeueYoshii2021}, such as WSe$_2$/black phosphorene heterostructure~\cite{AkamatsuIdeueYoshii2021} and twisted TMD heterobilayers~\cite{Hu2023,HuLouie2023}.

In the present work, we theoretically study the shift current response in AB-AB and AB-BA stacked TDBG.
This is motivated by the fact that TDBG is composed of AB-stacked BLG, which has broken $C_{2z}$ symmetry, allowing for a finite in-plane shift current response without the need for any symmetry-breaking unlike TBG.
Furthermore, the band structure of TDBG is highly tuneable through the application of a vertical bias voltage, giving an extra degree of freedom to investigate the behaviour of the shift current.
The fact that the two variants of TDBG exhibit similar band structures with distinct topologies ~\cite{Koshino2019,ChebroluJung2019,LiuDai2019} offers a platform to study how this contrast influences the shift current response.
A previous study investigated the shift current response in AB-AB and AB-BA stacked TDBG at a fixed twist angle \cite{LiuDai2020}.
In this work, we present a systematic analysis of the shift current in TDBG over a range of twist angles, Fermi levels, and vertical bias voltages --- parameter dependencies that were not addressed in the earlier study.
We find that the twist angle plays an important role of enhancing the signal strength as the twist angle is reduced, owing from the decrease in the typical size of the energy gaps.
We also place particular emphasis on the role of stacking configuration.
Notably, we discover a systematic sign reversal in the shift current between the AB-AB and AB-BA variants under large vertical bias.
We further provide a qualitative explanation for this sign reversal by analysing the shift current response in AB-stacked bilayer graphene.

The present paper is organised as follows.
In Section~\ref{sec:model}, we introduce the effective continuum Hamiltonian for TDBG and the theoretical expression of the shift current.
Then, in Section~\ref{sec:results}, we present our results of the shift current response in TDBG, placing particular emphasis on the effect of the twist angle and vertical bias, and provide a discussion on the interpretation on the results in Section~\ref{sec:blg}. Finally, we will conclude the present paper in Section~\ref{sec:conclusion}.

\section{Model and Methods}\label{sec:model}

\subsection{Twisted double bilayer graphene}\label{subsec:hamiltonian}
\begin{figure}[t]
    \centering
    \includegraphics[width=8.5cm]{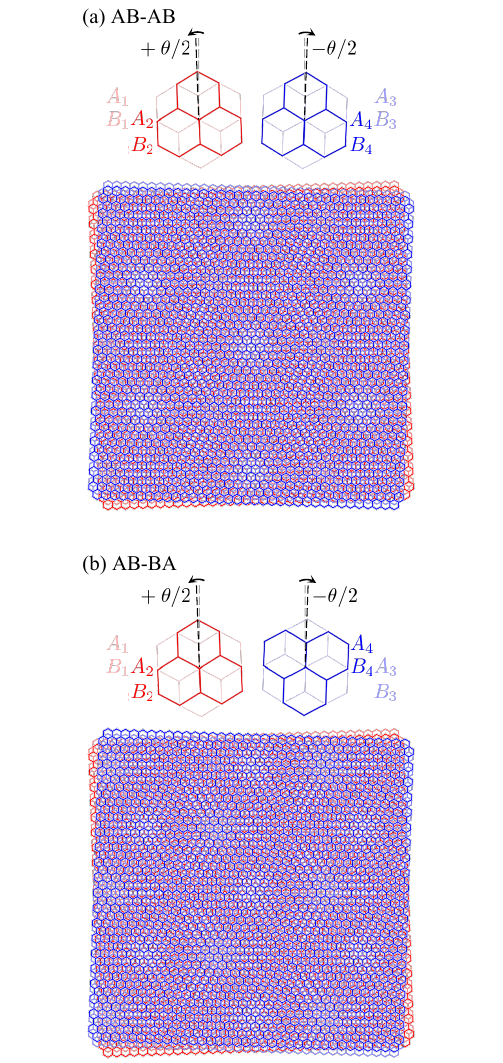}
    \caption{(a) We show a schematic diagram of AB-stacked BLG and the stacking configuration in AB-AB stacked TDBG on the top. The corresponding moir\'{e} pattern is shown in the bottom. In (b), we shows the same diagrams, but for AB-BA stacked TDBG, where the blue bilayer gains a $180^\circ$ offset.}
    \label{fig:3-structure}
\end{figure}

The primitive lattice vectors of the graphene lattice are chosen to be $\vec{a}_1=a(1,0)$ and $\vec{a}_2=a(1/2,\sqrt{3}/2)$, where $a=$~0.246~nm is the lattice constant.
Then, the corresponding reciprocal lattice vectors are $\vec{b}_1=4\pi/\sqrt{3}a(\sqrt{3}/2,-1/2)$ and $\vec{b}_2=4\pi/\sqrt{3}a(0,1)$.
AB-stacked bilayer graphene (BLG), which can be prepared by stacking two graphene sheets with the A and B sites of the upper and lower layers aligned, shares the same primitive lattice vectors with monolayer graphene, thus, the Brillouin zones are also common.
A schematic diagram is shown on the left side of Fig.~\ref{fig:3-structure} (a), where there are four atomic sites labelled $A_1$, $B_1$, $A_2$ and $B_2$.
Sites ($B_1,\,A_2$), which are vertically aligned, are referred to as the dimerised sites and experience an energy offset of $\Delta'=$~0.050~eV.

Twisted double bilayer graphene (TDBG) is a system where two sheets of AB-stacked BLG are stacked and twisted with a relative twist angle, as shown in Fig.~\ref{fig:3-structure} (a), (b).
To fabricate a sample of TDBG, we first stack the two BLG without twisting.
There are two ways to perform this operation: one is to simply stack the two BLG sheets with the same orientation and the other is to perform a $180^\circ$ rotation about the $z$-axis on one sheet and then stacking the two.
Then, a relative twist angle between the two BLG sheets is introduced to form a sample of TDBG.
Here, the first variant is known as AB-AB stacking, while the second is known as AB-BA stacking.
Their respective structures are shown in Fig.~\ref{fig:3-structure} (a) and (b), and the difference in their moir\'{e} patterns can be seen.
Hence, the extra $180^\circ$ rotation leads to a different environment at the twist interface between the two stacking configurations, resulting in different band topologies~\cite{Koshino2019}.
It should also be noted that the AB-AB variant has the in-plane three-fold rotational symmetry $C_{3z}$ inherited from monolayer graphene and an out-of-plane two-fold rotational symmetry along $C_{2x}$ along the $x$-axis.
Similarly, the AB-BA variant has $C_{3z}$ and $C_{2y}$ rotational symmetries.

In cases when the moir\'{e} lattice constant is much greater than the graphene lattice constant, \textit{i.e.} when the twist angle is small, we can construct an effective continuum Hamiltonian for TDBG around the $K_\pm$ valleys. When the lower bilayer ($l=1$) and the upper bilayer ($l=2$) are rotated by an angle $\mp\theta/2$, the reciprocal lattice vectors in the respective bilayers are given as $\vec{b}^{(l)}_i = R\left[(-1)^l\theta/2\right]\vec{b}_i$, where $R(\theta)$ is the two dimensional rotation matrix. This allows us to define the moir\'{e} reciprocal lattice vectors $\vec{G}^{\rm{M}}_i=\vec{b}^{(1)}_i-\vec{b}^{(2)}_i$.
 
The Hamiltonians for AB-AB and AB-BA TDBG are written as
\begin{align}
    \begin{split}
        H_{\rm{AB\text{-}AB}} &= 
        \begin{pmatrix}
            H(\vec{k}_1) & g^\dagger(\vec{k}_1) & &
            \\
            g(\vec{k}_1) & H'(\vec{k}_1) & U^\dagger(\vec{r}) &
            \\
            & U(\vec{r}) & H(\vec{k}_2) & g^\dagger(\vec{k}_2)
            \\
            & & g(\vec{k}_2) & H'(\vec{k}_2)
        \end{pmatrix} + V,
        \\
        H_{\rm{AB\text{-}BA}} &= 
        \begin{pmatrix}
            H(\vec{k}_1) & g^\dagger(\vec{k}_1) & &
            \\
            g(\vec{k}_1) & H'(\vec{k}_1) & U^\dagger(\vec{r}) &
            \\
            & U(\vec{r}) & H'(\vec{k}_2) & g(\vec{k}_2)
            \\
            & & g^\dagger(\vec{k}_2) & H(\vec{k}_2)
        \end{pmatrix} + V,
    \end{split}
    \label{eq:tdbghamiltonian}
\end{align}
where we have defined
\begin{align}
    \begin{split}
        H(\vec{k}) &=
        \begin{pmatrix}
            0 & -\hbar v_0k_-
            \\
            -\hbar v_0k_+ & \Delta'
        \end{pmatrix},
        \\
        H'(\vec{k}) &=
        \begin{pmatrix}
            \Delta' & -\hbar v_0k_-
            \\
            -\hbar v_0k_+ & 0
        \end{pmatrix},
        \\
        g(\vec{k}) &=
        \begin{pmatrix}
            \hbar v_4k_+ & \gamma_1
            \\
            \hbar v_3k_- & \hbar v_4k_+
        \end{pmatrix}.
    \end{split}
\end{align}
The velocities are defined as $v_i= \sqrt{3}a\abs{\gamma_i}/(2\hbar)$, $\gamma_i$ being hopping integrals, and $k_{\pm}=\xi k_x\pm ik_y$, $\xi$ being the valley index $K_\xi$.
Namely, $\gamma_0=-2.4657$~eV is the intralayer nearest neighbouring hopping, $\gamma_1=$~0.4~eV is the hopping between the dimerised sites, $\gamma_3=$~0.32~eV and $\gamma_4=$~0.044~eV are the diagonal hopping between the ($A_1$,\,$B_2$) and ($B_1$,\,$B_2$) sites, respectively.
Also, $\vec{k}_l=R\left[(-1)^l\theta/2\right](\vec{k}-\vec{K}^{(l)}_\xi)$ and $\vec{K}^{(l)}_\xi$ are the $\vec{K}_\xi$ valleys in the respective bilayers.
By inspecting the Hamiltoanians, we can see that the top left block corresponds to the $4\times4$ Hamiltonian for AB-stacked BLG. We have the same Hamiltonian in the bottom right block for $H_{\rm{AB\text{-}AB}}$, however, for $H_{\rm{AB\text{-}BA}}$, we can see that the pairs $[H(\vec{k}_2), H'(\vec{k}_2)]$ and $[g(\vec{k}_2), g^\dagger(\vec{k}_2)]$ are interchanged, reflecting the difference in stacking configuration.
The interlayer moir\'{e} potential $U(\vec{r})$ is given by
\begin{align}
    U(\vec{r}) = 
    \begin{pmatrix}
        u & u'
        \\
        u' & u
    \end{pmatrix}
    & +
    \begin{pmatrix}
        u & u'\omega^{-\xi}
        \\
        u'\omega^\xi & u
    \end{pmatrix}
    e^{i\xi\vec{G}^{\rm{M}}_1\vdot\vec{r}}
    \\
    \nonumber
    & +
    \begin{pmatrix}
        u & u'\omega^\xi
        \\
        u'\omega^{-\xi} & u
    \end{pmatrix}
    e^{i\xi\left(\vec{G}^{\rm{M}}_1+\vec{G}^{\rm{M}}_2\right)\vdot\vec{r}},
\end{align}
where $u=$~0.0797~eV, $u'=$~0.0975~eV~\cite{Koshino2019,KoshinoFu2018} and $\omega=\exp(2\pi i/3)$ is the cube root of unity. The final term $V$ is the diagonal matrix modelling the vertical bias voltage
\begin{equation}
    V = 
    \begin{pmatrix}
        \frac{3}{2}\Delta\,\mathbb{I} & & &
        \\
        & \frac{1}{2}\Delta\,\mathbb{I} & &
        \\
        & & -\frac{1}{2}\Delta\,\mathbb{I} &
        \\
        & & & -\frac{3}{2}\Delta\,\mathbb{I}
    \end{pmatrix}.
\end{equation}
We note that a finite $\Delta$ breaks the $C_{2x}$/$C_{2y}$ rotational symmetries in the AB-AB/AB-BA stacked TDBG.
In this system, the twist angle $\theta$ and the vertical bias voltage $\Delta$ are the parameters that can be modified experimentally.

The numerical evaluation of Hamiltonians in Eq.~\eqref{eq:tdbghamiltonian} involves expanding $U(\vec{r})$ in term of the Fourier components with respect to $\vec{G}=n_1\vec{G}^{\rm{M}}_1+n_2\vec{G}^{\rm{M}}_2$, with integers $n_1$ and $n_2$.
This Fourier expansion entails the introduction of a cutoff in $\vec{k}$ space, and we have extended the cutoff until convergence was achieved, corresponding to $\norm{\vec{k}}\leq 5G^{\rm{M}}$.

\subsection{Shift current}\label{subsec:shift}
When two rays of light, $E_{\alpha}(\omega_\alpha)$ and $E_{\beta}(\omega_\beta)$ polarised in the $\alpha,~\beta$ direction with frequency $\omega_\alpha,~\omega_\beta$ respectively, are shone to a sample, we expect a current density in the $\mu$ direction $j_\mu$ resulting from a second order nonlinear optical response (NLOR).
Here, $\mu,\alpha,\beta = x,y$ in the present 2D set-up. Adopting the Einstein summation convention over repeated indices, the second order NLOR is expressed as
\begin{equation}
    j_\mu(\omega_\Sigma) = \sigma^\mu_{\alpha\beta}(\omega_\Sigma;\omega_\alpha,\omega_\beta)\,E_{\alpha}(\omega_\alpha)E_{\beta}(\omega_\beta),
    \label{eq:secondorderdef}
\end{equation}
where $\sigma^\mu_{\alpha\beta}$ is the second order NLOR conductivity tensor and $\omega_\Sigma = \omega_\alpha+\omega_\beta$.
It should be noted that $\sigma^\mu_{\alpha\beta}$ will be finite only when the system lacks inversion symmetry~\cite{vonBaltzKraut1981,Sipe2000}, which can be checked by inspecting the inversion symmetry of both sides of Eq.~\eqref{eq:secondorderdef}.

The shift current corresponds to the case where $\omega_\alpha = -\omega_\beta = \omega$, resulting in a dc response $j_\mu(0)$.
If we choose to focus on the case where the incoming light is of a single polarisation $\alpha$, the shift current conductivity in 2D is given by the following expression~\cite{Sipe2000,ParkerMorimoto2019}
\begin{equation}
    \sigma^\mu_{\alpha\alpha}(\omega) = \frac{2\pi q^3}{\hbar\omega^2}\int\frac{d^2\vec{k}}{(2\pi)^2}\sum_{a,b}f_{ab}\abs{v^\alpha_{ba}}^2R^{\mu(\alpha)}_{ba}\delta(\hbar\omega-\varepsilon_{ba}).
    \label{eq:scoriginal}
\end{equation}
Here, $q$ is the charge of the carrier, $\varepsilon_{ba} = \varepsilon_{b} - \varepsilon_{a}$, where $\varepsilon_{a}$ is the eigenenergy of the Bloch state $\ket{a}$, $f_{ab} = f(\varepsilon_a) - f(\varepsilon_b)$, where $f(\varepsilon)$ is the Fermi-Dirac occupation function, $v^\alpha = \hbar^{-1}\partial_\alpha H$ is the velocity operator, where $\partial_\alpha$ is used as a shorthand for $\partial/\partial k_\alpha$.
We also introduce the shift vector $R^{\mu(\alpha)}_{ba}(\vec{k}) = A^\mu_{bb} - A^\mu_{aa} - \partial_\mu\varphi^{(\alpha)}_{ba}$, defined using the intraband Berry connection $A^\alpha_{aa} = i\mel{a}{\partial_\alpha}{a}$ and the phase of the velocity matrix element $\varphi^{(\alpha)}_{ba} = \arg(v^\alpha_{ba})$.

A close inspection of Eq.~\eqref{eq:scoriginal} allows us to deduce a couple of key features of the shift current.
First, $f_{ab}\,\delta(\hbar\omega-\varepsilon_{ba})$ is the joint density of states (JDoS) which counts the number of states that are separated in energy by $\hbar\omega$, and $\abs{v^\alpha_{ba}}^2$ is the dipole matrix element which dictates the transition rules at each $k$-point.
Thus, we can conclude that $a$ and $b$ respectively labels the initial and final state of the optical transition and that the magnitude of the current is proportional to the number of states available for the transition.
Second, the shift current is characterised by the shift vector $R^{\mu(\alpha)}_{ba}$.
It is expressed as the difference between the intraband Berry connections $A^\mu_{nn}$ of the initial and final bands of the optical transition, and we can understand this as being the difference in the centre-of-mass coordinate of the electron wavepacket in the initial and the final band~\cite{Wannier1937,KingSmithVanderbilt1993,Resta1994}.
The final $\partial_\mu\varphi^{(\alpha)}_{ba}$ term is present to ensure that $R^{\mu(\alpha)}_{ba}$ is gauge invariant. 

In the present paper, we adopt the velocity gauge for numerical evaluations.
In this gauge, the electromagnetic field is introduced via the minimal coupling approach~\cite{ParkerMorimoto2019} and is the preferred method when the velocity operators are known; the velocity operators can be easily calculated from Eq.~\eqref{eq:tdbghamiltonian}.
Furthermore, we only work with Hamiltonians which are linear in $\vec{k}$, which allows us to rewrite Eq.~\eqref{eq:scoriginal} as~\cite{Chaudhary2022}
\begin{align}
    \sigma^\mu_{\alpha\alpha}(\omega) &= -\frac{\hbar^2 e^3}{2\pi}\int d^2\vec{k} \sum_{a,b}\frac{f_{ab}}{\varepsilon_{ba}^2}\;\times \nonumber
    \\
    &\Im\left[\sum_{c\neq a}\frac{v^\mu_{ac}v^\alpha_{cb}v^\alpha_{ba}}{\varepsilon_{ac}} + \sum_{c\neq b}\frac{v^\mu_{cb}v^\alpha_{ba}v^\alpha_{ac}}{\varepsilon_{bc}}\right]\delta(\hbar\omega-\varepsilon_{ba}),
    \label{eq:scnumerical}
\end{align}
where we have set $q=-e$, where $e$ is the elemental charge.
Here, we can view the states $a$ and $b$ as the initial and final states of the real optical transition, while state $c$ corresponds to some intermediate state that is reached through virtual transitions.
The numerical evaluation of the shift current response in the present work was performed using this expression.
We replace the integral over the moir\'{e} Brillouin zone by a summation over a mesh of $50\times50$ and employ a Lorentzian function $\eta/[(\hbar\omega-\varepsilon_{ba})^2 + \eta^2]$ with a broadening $\eta=$~1~meV in place of $\delta(\hbar\omega-\varepsilon_{ba})$.
We note that this replacement may lead to a finite signal as $\omega \to 0$, where the response should strictly vanish, when the gap size $\varepsilon_{ba}$ becomes comparable to the broadening parameter $\eta$.
This unphysical behaviour is an artifact of the calculation and does not affect the results for $\omega \gtrsim \eta$, where the approximation remains valid.

In applying Eq.~\eqref{eq:scnumerical} to our system, we can perform a simple symmetry analysis on the $\sigma^\mu_{\alpha\beta}$ tensor to find its independent components.
To do so, we return to Eq.~\eqref{eq:secondorderdef} and consider a transformation from coordinates $\vec{r}^\prime=(x^\prime,y^\prime)$ to a new set of coordinates $\vec{r}=(x,y)$ given by a rotation matrix $\vec{r}^\prime=R(\psi)\,\vec{r}$.
In this new frame, we can rewrite Eq.~\eqref{eq:secondorderdef} as
\begin{align}
    j_{\mu^\prime} &= \sigma^{\mu^\prime}_{\alpha^\prime\beta^\prime}\;E_{\alpha^\prime}E_{\beta^\prime} \nonumber
    \\
    R_{\mu^\prime\mu}j_{\mu} &= \sigma^{\mu^\prime}_{\alpha^\prime\beta^\prime}\;R_{\alpha^\prime\alpha}E_\alpha\;R_{\beta^\prime\beta}E_\beta \nonumber
    \\
    j_\mu &= \left(R^{-1}_{\mu\mu^\prime}\sigma^{\mu^\prime}_{\alpha^\prime\beta^\prime}R_{\alpha^\prime\alpha}R_{\beta^\prime\beta}\right)E_\alpha E_\beta,
    \label{eq:transformation}
\end{align}
leading us to the following transformation law
\begin{equation}
    \sigma^\mu_{\alpha\beta}=R^{-1}_{\mu\mu^\prime}\sigma^{\mu^\prime}_{\alpha^\prime\beta^\prime}R_{\alpha^\prime\alpha}R_{\beta^\prime\beta}.
    \label{eq:transflaw}
\end{equation}
Imposing $C_{3z}$ symmetry, in which we set $\sigma^\mu_{\alpha\beta}=\sigma^{\mu^\prime}_{\alpha^\prime\beta^\prime}$ with $\psi=2\pi/3$, we find~\cite{LiuDai2020}
\begin{align}
    \begin{split}
        \sigma^x_{xx} &= -\sigma^y_{xy} = -\sigma^y_{yx} = -\sigma^x_{yy},
        \\
        \sigma^y_{yy} &= -\sigma^x_{xy} = -\sigma^x_{yx} = -\sigma^y_{xx}.
    \end{split}
    \label{eq:symmetry}
\end{align}
Therefore, $\sigma^\mu_{\alpha\beta}$ only contains two independent components, $\sigma^x_{xx}$ and $\sigma^y_{yy}$.
Since any signal can be expressed in terms of $\sigma^x_{xx}$ and $\sigma^y_{yy}$, in the subsequent sections, we solely focus on these two components.
We further note that in the case where the vertical bias voltage is absent, AB-AB (AB-BA) configuration recovers the out-of-plane $C_{2x}$ ($C_{2y}$) rotational symmetry~\cite{Koshino2019}, further rendering $\sigma^y_{yy}$~=~0 ($\sigma^x_{xx}$~=~0).

We finally note that in the present system, the two valleys $K_{\pm}$ are connected by time-reversal symmetry.
This ensures that the contributions to the shift current from each valley are equal~\cite{KaplanHolder2022,Chaudhary2022} and the following results have taken this degeneracy into account.

\section{Results}\label{sec:results}
In this section, we present the numerical results on the shift current response in AB-AB and AB-BA twisted double bilayer graphene (TDBG) at various twist angles $\theta$ and vertical bias voltages $\Delta$.
We begin by studying how the shift current evolves as we vary $\theta$, specifically focussing on the intrinsic signal at $\Delta=$~0.
We then investigate the effect of $\Delta$ on the response, fixing $\theta=0.8^\circ$, and compare the resulting signals in the two variants of TDBG.

\subsection{Evolution with twist angle}\label{subsec:theta}

\begin{figure*}[t]
    \centering
    \includegraphics[width=17cm]{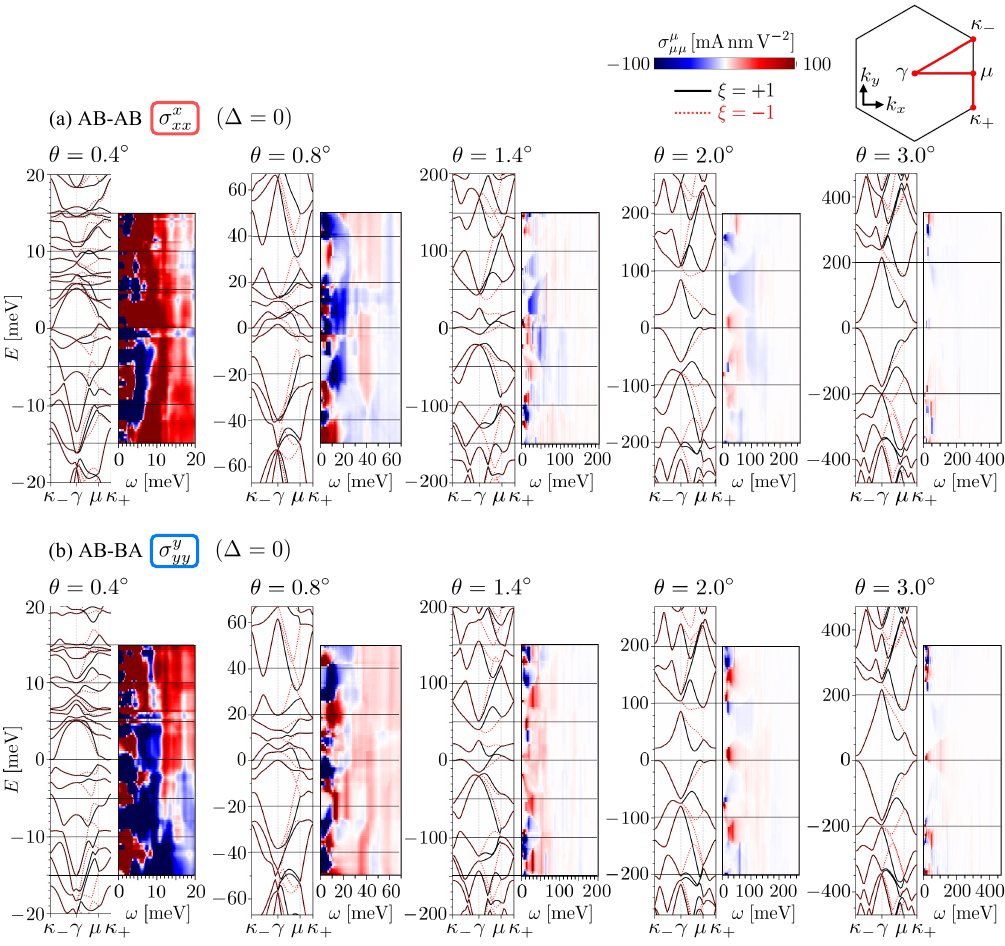}
    \caption{(a) The band structures of AB-AB stacked TDBG and density plots of $\sigma^x_{xx}(\omega;E_{\rm{F}})$ signal for various twist angles. Here, we used $\theta=0.4^\circ, 0.8^\circ, 1.4^\circ, 2.0^\circ$ and $3.0^\circ$. (b) The corresponding plots for the $\sigma^y_{yy}(\omega;E_{\rm{F}})$ signal in the AB-BA variant. We note that the $\sigma^y_{yy}$/$\sigma^x_{xx}$ component vanishes due to the $C_{2x}$/$C_{2y}$ symmetry in the AB-AB/AB-BA variant. The moir\'{e} Brillouin zone in shown in the top right corner.}
    \label{fig:3-tdbg04-40}
\end{figure*}

We show the intrinsic shift current responses for $\theta=0.4^\circ, 0.8^\circ, 1.4^\circ, 2.0^\circ$ and $3.0^\circ$ in Fig.~\ref{fig:3-tdbg04-40}.
In each panel of Fig.~\ref{fig:3-tdbg04-40} (a)/(b), we show the band structures and the density plots of $\sigma^\mu_{\mu\mu}(\omega;E_{\rm{F}})$ for AB-AB/AB-BA TDBG for each value of $\theta$.
We note that the figures only show the non-vanishing component,
$\sigma^x_{xx}$ for AB-AB and $\sigma^y_{yy}$ for AB-BA;
the omitted components are suppressed by the out-of-plane rotational symmetry as argued in the previous section.
The shift current signals $\sigma^\mu_{\mu\mu}(\omega;E_{\rm{F}})$ are given as colour density plots on the right side of the panels with frequency $\omega$ on the horizontal and Fermi level $E_{\rm{F}}$ on the vertical axis.
Therefore, each horizontal strip of the density plot corresponds to a $\sigma^\mu_{\mu\mu}$~-~$\omega$ plot for a certain value of $E_{\rm{F}}$.
We note that the scale of the vertical axis is adjusted to match that of the band structure plot and the colour scale of the density plots is shared between all plots.


In the band structures, we observe a gradual formation of moir\'{e} subbands as the twist angle decreases, where the hyperbolic bands inherited from AB-stacked bilayer graphene are reconstructed into a series of flat bands.
Then, inspecting the $\sigma^\mu_{\mu\mu}(\omega;E_{\rm{F}})$ density plots, we find that there is an overall trend of the signal strength to significantly enhance as $\theta$ is decreased. There are a couple of reasons behind this.
First, the typical energy gap becomes smaller at lower twist angles, amplifying the response through the $1/\omega^2$ prefactor in Eq.~\eqref{eq:scoriginal}.
Second, the moir\'{e} coupling weakens in the large-angle limit. In this regime, the low-energy electronic structure approaches that of decoupled AB-stacked bilayer graphene, which possesses inversion symmetry and therefore suppresses the shift current. As a result, the signal is reduced at larger twist angles, where the properties of the individual bilayers dominate.


We also notice that, at small angles $\theta=0.4^\circ,0.8^\circ$, the low frequency signal is not of a single sign but a series of sign flips is observed as the Fermi level is swept.
We can understand this in the following manner.
The contribution to the low frequency signal is due to the transitions between neighbouring flat bands.
As we change $E_{\rm{F}}$ and the band edge is crossed, the contribution to the signal now comes from a different band transition, leading to a sign flip in the shift current.
This also explains why we do not see such abrupt sign changes in the high frequency regime where multiple transitions contribute to the signal.
When $\theta$ is increased, we begin to lose this feature as the bands become more dispersive.
Since this phenomenon is due to the formation of a series of moir\'{e} flat bands, the series of sign flips in the signal is a unique feature of moir\'{e} systems.

\subsection{Effect of vertical bias voltage}\label{subsec:delta}

\begin{figure*}[t]
    \centering
    \includegraphics[width=17cm]{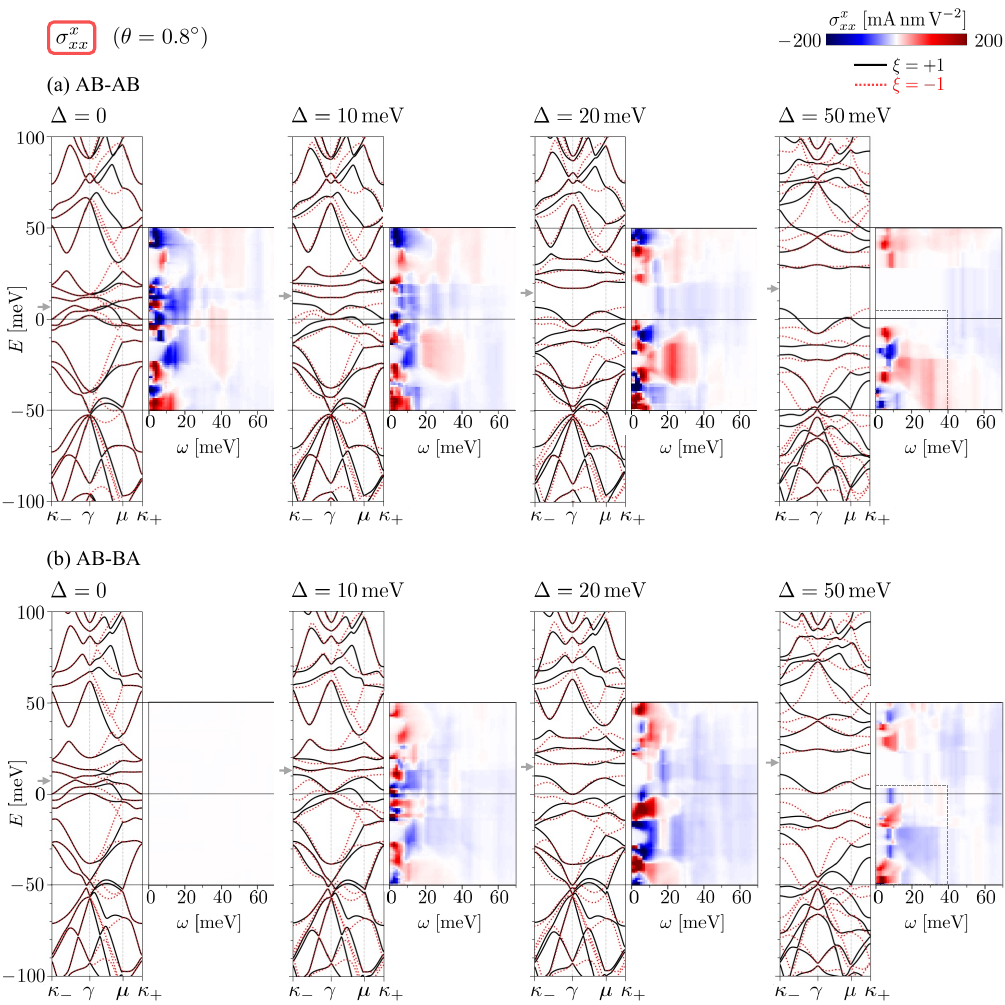}
    \caption{The band structures and $\sigma^x_{xx}(\omega;E_{\rm{F}})$ density plots of (a) AB-AB and (b) AB-BA stacked TDBG for various strengths of the vertical bias voltage $\Delta$. Here, we used $\Delta=$~0, 10, 20 and 50~meV. The grey arrows indicate the respective charge neutral gaps.}
    \label{fig:3-tdbg0800-50xxx}
\end{figure*}

\begin{figure*}[t]
    \centering
    \includegraphics[width=17cm]{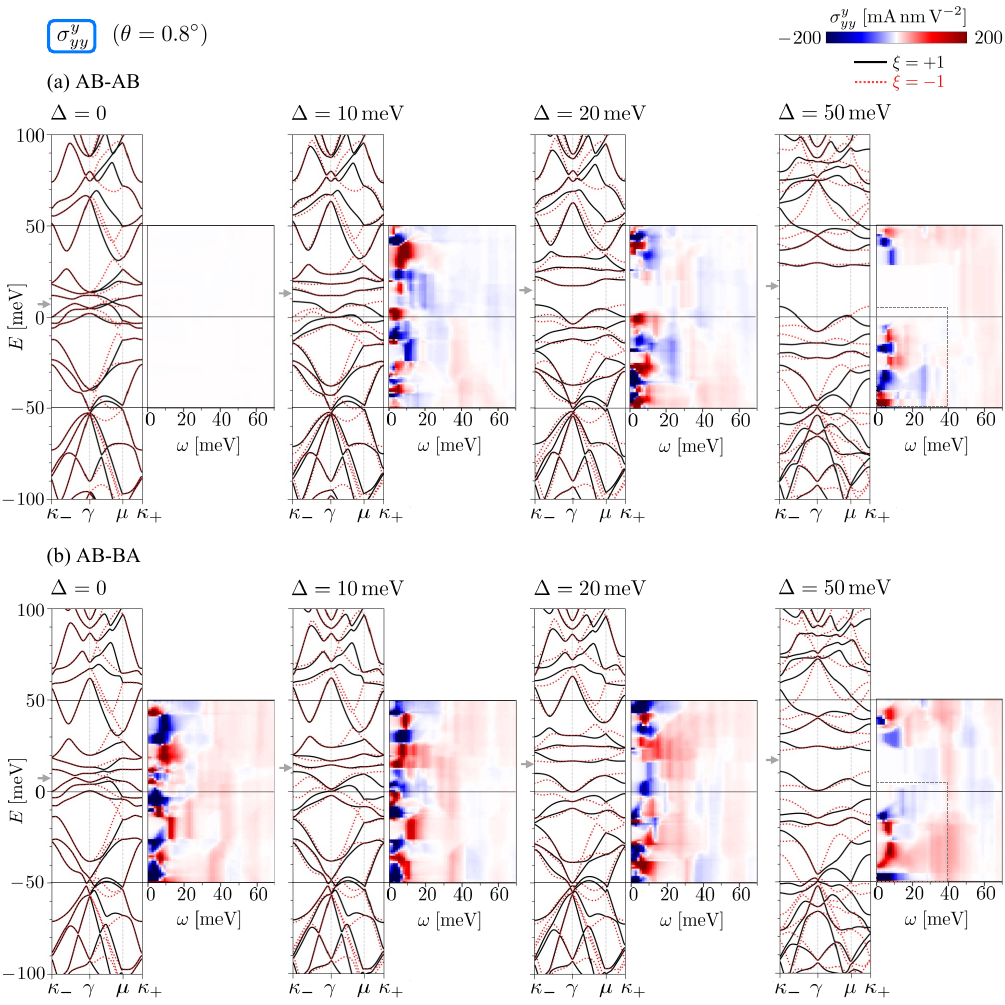}
    \caption{The band structures and $\sigma^y_{yy}(\omega;E_{\rm{F}})$ density plots of (a) AB-AB and (b) AB-BA stacked TDBG for various strengths of the vertical bias voltage $\Delta$. Here, we used $\Delta=$~0, 10, 20 and 50~meV. The grey arrows indicate the respective charge neutral gaps.}
    \label{fig:3-tdbg0800-50yyy}
\end{figure*}

We now move our attention to the evolution of the response as a result of the application of $\Delta$.
We begin with the $\sigma^x_{xx}(\omega;E_{\rm{F}})$ response at values of $\Delta=$~10, 20 and 50~meV and show the band structures of AB-AB and AB-BA stacked TDBG in Fig.~\ref{fig:3-tdbg0800-50xxx} (a) and (b) respectively.

In the band structures, as $\Delta$ is increased, we find that the bands become well separated and a wide gap opens at the charge neutrality point (CNP).
Focussing on the AB-AB configuration, where $\sigma^x_{xx}$ is always non-zero, as $\Delta$ is increased, we see that the overall size of the signal is suppressed.
This is due to the fact that the bias voltage increases the gap size, which in turn suppresses the signal via the $1/\omega^2$ factor.
On the other hand, in the AB-BA variant, as soon as the vertical bias voltage begins to break the $C_{2y}$ symmetry, we observe a finite response.
From here, the signal continues to grow in magnitude until $\Delta\sim$~10~meV, from which the signal follows the same pattern with the $1/\omega^2$ suppression for stronger $\Delta$.

We further notice that there is a large insulating region, represented in white, near the CNP as $\Delta$ increases. 
This behaviour can be simply inferred from the large band gap that opens at the CNP.
We see similar trends in the $\sigma^y_{yy}$ response show in in Fig.~\ref{fig:3-tdbg0800-50yyy} (a) and (b).


\subsection{Comparison between AB-AB and AB-BA stacked TDBG}\label{subsec:comparison}


We find that there is a visible relationship between the signals of the two variants when $\Delta$ is large and the bands are well separated.
To allow for a systematic comparison of the shift current response between the two stacking configurations, we consider the case where a finite vertical bias voltage of $\Delta=$~50~meV is applied.
On the rightmost side of Fig.~\ref{fig:3-tdbg0800-50xxx}, we show the band structure and the corresponding $\sigma^x_{xx}$ plots for (a) AB-AB and (b) AB-BA stacked TDBG.

\begin{figure}[t]
    \centering
    \includegraphics[width=8.5cm]{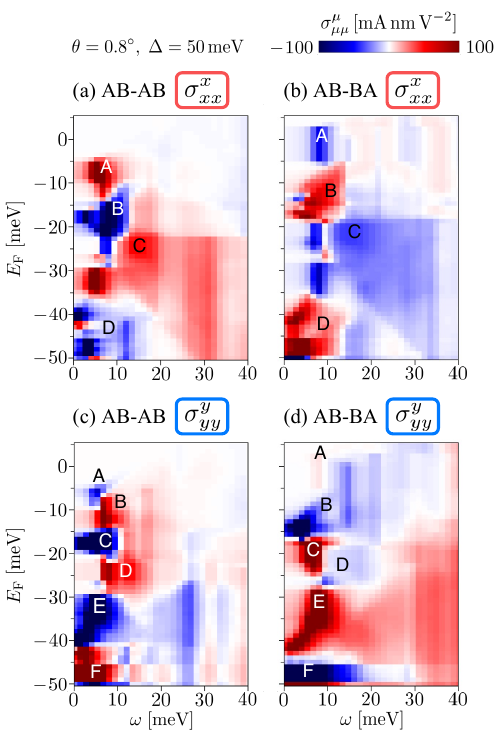}
    \caption{A close-up view of the $\sigma^x_{xx}$/$\sigma^y_{yy}$ responses in (a)/(c) AB-AB and (b)/(d) AB-BA TDBG shown in Fig.~\ref{fig:3-tdbg0800-50xxx} and \ref{fig:3-tdbg0800-50yyy}. We have labelled regions where a sign reversal between the two stacking configurations can be seen.}
    \label{fig:3-zoomin}
\end{figure}

If we begin to compare the density plots between the two stacking configurations, we notice that there is a general positive/negative correlation in the sign of the signal above/below the CNP.
The negative correlation is illustrated in Fig.~\ref{fig:3-zoomin} where we show a close-up view of the four density plots below the CNP, indicated by the grey dashed boxes in Fig.~\ref{fig:3-tdbg0800-50xxx} and \ref{fig:3-tdbg0800-50yyy}.
Focussing first on the $\sigma^x_{xx}$ response shown in Fig.~\ref{fig:3-zoomin} (a) and (b), we have labelled four regions from A to D where a sign reversal can be seen.
For example, in the AB-AB variant, region A has a positive signal, while the corresponding signal in the AB-BA variant is negative.
A similar pattern is seen for the other three regions.
Likewise, in Fig.~\ref{fig:3-zoomin} (c) and (d), where show the density plots for $\sigma^y_{yy}$, we find that regions with the same label have opposite signs.
We note that this positive/negative correlation is overshadowed as we turn down $\Delta$ and the bands begin to cluster.
We also give the shift current density maps for $\theta=2.0^\circ$ in Appendix.~\ref{appx:extracalc}, and show that this correlation is seen at other twist angles.

\section{Understanding sign reversal in TDBG shift current}\label{sec:blg}

In order to understand the sign reversal seen in the AB-AB and AB-BA stacked TDBG, we consider the shift current response in AB-stacked bilayer graphene (BLG).
We note that BLG is intrinsically centrosymmetric, hence, requires some external symmetry breaking for a finite response.
In the present case, we apply a vertical bias voltage of $\Delta=$~50~meV to match the value of $\Delta$ in the previous section.
It should also be noted that BLG has $M_x$ mirror symmetry across the $y$-$z$ plane, which renders $\sigma^x_{xx}$ to zero.
Therefore, it will suffice to study $\sigma^y_{yy}$ to describe its shift current response.

The continuum Hamiltonian for BLG around the $K_\pm$ valleys is given by the top left $4\times4$ matrix in Eq.~\eqref{eq:tdbghamiltonian},
\begin{equation}
        H_{\rm{AB}} = 
    \begin{pmatrix}
        H(\vec{k}) & g^\dagger(\vec{k})
        \\
        g(\vec{k}) & H'(\vec{k})
    \end{pmatrix}
    +V_{\rm{AB}},
    \label{eq:abhamiltonian}
\end{equation}
where $V_{\rm{AB}}$ represents the vertical bias voltage
\begin{equation}
       V_{\rm{AB}} = 
    \begin{pmatrix}
        \frac{1}{2}\Delta\,\mathbb{I} & 
        \\
        & -\frac{1}{2}\Delta\,\mathbb{I}
    \end{pmatrix}.
    \label{eq:blgv}
\end{equation}

\begin{figure}[t]
    \centering
    \includegraphics[width=8.5cm]{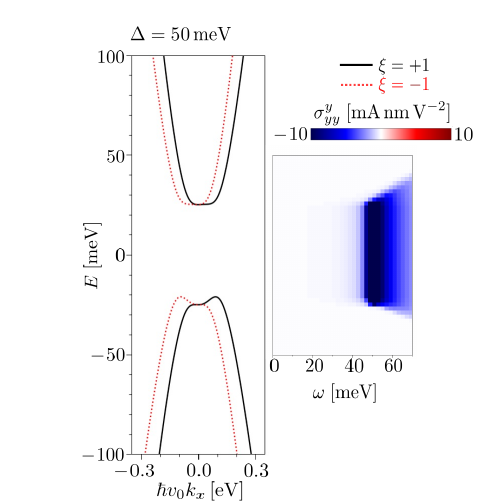}
    \caption{The band structure of AB-stacked BLG from the $K_+$ ($K_-$) is plotted in the solid black (dashed red) lines in the left panel; a vertical bias voltage of 50~meV is applied across the layers. The shift current $\sigma^y_{yy}(\omega;E_{\rm{F}})$ is plotted as a density plot in the right panel.}
    \label{fig:3-abblg}
\end{figure}

Using this Hamiltonian, we show the band structure (left panel) and the corresponding $\sigma^y_{yy}(\omega;E_{\rm{F}})$ response (right panel) of BLG in Fig.~\ref{fig:3-abblg}.
The band structure computed from the Hamiltonian in Eq.~\eqref{eq:abhamiltonian} shows a slice through $k_x$ at $k_y = 0$ around the $K_+$ ($K_-$) point, plotted in the black solid (red dashed) line.
The vertical bias voltage allows for a gap-opening of around 50~meV.
On the right panel, we show the shift current response as a colour density plot with frequency $\omega$ on the horizontal and Fermi level $E_{\rm{F}}$ on the vertical axis.
The numerical evaluation of Eq.~\eqref{eq:scnumerical} was performed with a broadening of $\eta=$~1~meV.
We can initially see that there is very little or no response at frequencies below 50~meV, which is a reflection of the fact that there are no possible optical transitions in this frequency range due to the gap opened via the bias voltage.
As we move up to $\omega=$~50~meV, we start to see a large response when the Fermi level is set within the gap.
This sharp response corresponds to the optical transition from the band edge of the valence to that of the conduction band, where a large joint density of states (JDoS) is realised.
We also note that the size of the signal is in the range of 10~mA~nm~$\rm{V}^{-2}$, which is an order of magnitude smaller compared to the signal seen in TDBG.
This result suggests that the effect of the moir\'{e} reconstruction of the band structure indeed enhances the shift current response.
We also note that the sign of the signal will be reversed through a $180^\circ$ rotation of the entire system.

We next move on to the case where two AB-stacked BLG are stacked and twisted with no coupling at the twist interface, which we will refer to as uncoupled TDBG.
Since the two sheets of BLG are uncoupled, the shift current response of the total system is given by the sum of the responses from each BLG. 

\begin{figure}[t]
    \centering
    \includegraphics[width=8.5cm]{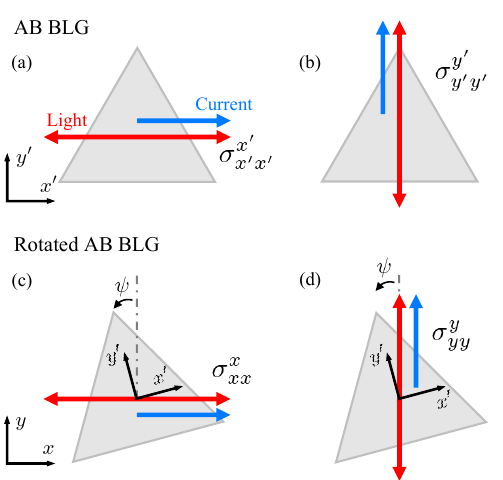}
    \caption{A schematic diagram of the set up of the shift current conductivities (a) $\sigma^{x^\prime}_{x^\prime x^\prime}$, (b) $\sigma^{y^\prime}_{y^\prime y^\prime}$ in the coordinates aligned with the BLG lattice orientation and (c) $\sigma^x_{xx}$, (d) $\sigma^y_{yy}$ in the fixed coordinates, where the BLG sheet is rotated through an angle $\psi$.}
    \label{fig:3-nonmoiresetup}
\end{figure}

The results of the shift current shown in Fig.~\ref{fig:3-abblg} are given with respect to coordinates aligned with the lattice orientation of the BLG, which we will denote using primed coordinates $\sigma^{y^\prime}_{y^\prime y^\prime}$.
However, once we begin to rotate the BLG sheets, we are in need to compute the components of the shift current, $\sigma^x_{xx}$ and $\sigma^y_{yy}$, with respect to the fixed coordinates denoted by unprimed coordinates. 
A schematic diagram of the set-up under consideration is shown in Fig.~\ref{fig:3-nonmoiresetup}. 
Using the transformation law in Eq.~\eqref{eq:transflaw}, the relationship between the conductivities are given as
\begin{align}
    \begin{split}
        \sigma^x_{xx}(\psi) &= \sigma^{y^\prime}_{y^\prime y^\prime}\sin(3\psi),
        \\
        \sigma^y_{yy}(\psi) &= \sigma^{y^\prime}_{y^\prime y^\prime}\cos(3\psi),
    \end{split}
    \label{eq:sigmatransformed}
\end{align}
where we have used the fact that $\sigma^{x^\prime}_{x^\prime x^\prime} = 0$.
The detail of the derivation is described in Appendix.~\ref{appx:transformation}.
We note that, now, both $\sigma^x_{xx}$ and $\sigma^y_{yy}$ are finite since $M_x$ symmetry is broken as the two BLG sheets are twisted relative to each other.

\begin{figure}[t]
    \centering
    \includegraphics[width=8.4cm]{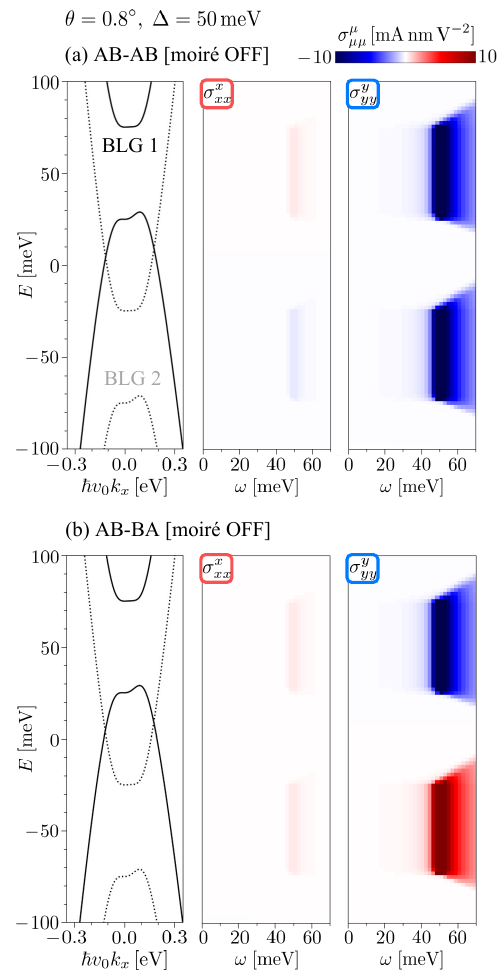}
    \caption{The band structure of (a) AB-AB and (b) AB-BA stacked uncoupled TDBG is shown in the left panels. The plot is around the $K_+$ point, the twist angle is set to $\theta=0.8^\circ$ and a vertical bias voltage of $\Delta=$~50~meV is applied. The contribution from the first (second) BLG is plotted in the solid (dotted) black lines. The corresponding moir\'{e} reciprocal lattice vector is also labelled. On the centre and right panels, we show the corresponding $\sigma^x_{xx}$ and $\sigma^y_{yy}$ plots, plotting frequency $\omega$ and Fermi energy $E_{\rm{F}}$ on the horizontal and vertical axis respectively.}
    \label{fig:3-nonmoire}
\end{figure}

Now, that we have an expression for $\sigma^x_{xx}$ and $\sigma^y_{yy}$, we will apply this to the uncoupled TDBG. Here, we use the parameters $\theta=0.8^\circ$ and $\Delta=$~50~meV.
In Fig.~\ref{fig:3-nonmoire}, we show the schematic band structures of (a) AB-AB and (b) AB-BA uncoupled TDBG at the $K_+$ point, which simply corresponds to having two copies of the band structure shown in Fig.~\ref{fig:3-abblg} with a shift in the Fermi energy.
On the right panels, we show the density plots for the shift current conductivities $\sigma^x_{xx}$ and $\sigma^y_{yy}$.
Fist, we note that the magnitudes of $\sigma^x_{xx}$ is much smaller than $\sigma^y_{yy}$, which stems from the fact that $\sigma^x_{xx}$ grows linearly with $\theta$ for small twist angles.



If we focus on the signs of the $\sigma^x_{xx}$ signal on the centre panels, we can see that there is a positive/negative correlation above/below the charge neutrality point between the two variants.
This can be understood from the fact that the relative orientation of the BLG sheets are aligned/antialigned in the AB-AB/AB-BA variant.
As the shift current is related to the polarisation of the material, the $180^\circ$ rotation of the second bilayer leads to a sign reversal of the signal.
The same effect is seen in the $\sigma^y_{yy}$ signal on the right panels.
This qualitatively explains the sign reversal that was seen in the TDBG calculation, and it reveals that this relationship is retained even after the effect of the moir\'{e} coupling.

\section{Conclusion}\label{sec:conclusion}

We have performed a theoretical investigation on the shift current response in AB-AB and AB-BA stacked twisted double bilayer graphene (TDBG).
We investigated the evolution of the TDBG shift current response as the twist angle $\theta$, Fermi level $E_{\rm{F}}$ and vertical bias voltage $\Delta$ were varied.
We found that there is a strong enhancement of the signal as $\theta$ is reduced, owing from the small gap size in the moir\'e subband structure.
We also found a series of sign flips in the low frequency regime as $E_{\rm{F}}$ is swept, which is a consequence of the formation of a series of moir\'{e} flat bands.


Concerning the effects of the vertical bias voltage, we find that the $\sigma^\mu_{\mu\mu}$ component, which vanishes at $\Delta = 0$, rapidly increases as a finite $\Delta$ is applied, breaking the out-of-plane rotational symmetry.
Importantly, we have found a positive/negative correlation in the sign of the signal between the AB-AB and AB-BA variants above/below the charge neutrality point at large values of the vertical bias voltage.
To understand the origin of the sign relationship between the two variants, we considered an uncoupled TDBG model with no moir\'{e} interaction at the twist interface.
The analysis of this model revealed that the sign reversal arises from the $180^\circ$ rotation of one of the bilayers.
Remarkably, this relative sign difference between the variants persists even after including the effects of moir\'{e}-induced band reconstruction.

\begin{acknowledgements}
This work was supported by JSPS KAKENHI Grants No. JP20K14415, No. JP20H01840, No. JP20H00127, No. JP21H05236, No. JP21H05232, JP24K06921, by JST CREST Grant No. JPMJCR20T3, and by JST SPRING, Grant No. JPMJSP2138, Japan.
\end{acknowledgements}

\appendix

\section{Deriving the expressions of the shift current response}\label{appx:derivation}

Here, we give an outline of the derivations of Eq.\eqref{eq:scoriginal} and Eq.\eqref{eq:scnumerical}.
Starting with the former, we refer to the general expression for the second order conductivity tensor given in Ref.~\cite{ParkerMorimoto2019} Eq.(43), which reads
\begin{widetext}
\begin{align}
    \sigma^\mu_{\alpha\beta}(\omega_\Sigma;\omega_\alpha,\omega_\beta) 
    &= -\frac{q^3}{\omega_\alpha\omega_\beta}\int\frac{d^2\vec{k}}{(2\pi)^2}\sum_{a,b,c} \left[\frac{1}{2}f_a\,v^{\mu\alpha\beta}_{aa} +f_{ab}\frac{v^{\mu\beta}_{ab}v^\alpha_{ba}}{i\omega_\alpha - \varepsilon_{ba}} + \frac{1}{2}f_{ab}\frac{v^\mu_{ab}v^{\alpha\beta}_{ba}}{i\omega_\Sigma - \varepsilon_{ba}} \right. \nonumber
    \\
    &\ \ \ \ \left. + \frac{v^\mu_{ac}v^\beta_{cb}v^\alpha_{ba}}{i\omega_\Sigma - \varepsilon_{ca}}\left(\frac{f_{ab}}{i\omega_\alpha - \varepsilon_{ba}}-\frac{f_{bc}}{i\omega_\beta - \varepsilon_{cb}}\right)\right] + [(\alpha,i\omega_\alpha) \leftrightarrow (\beta,i\omega_\beta)],
    \label{eq:parkermorimoto}
\end{align}
where $v^{\alpha_1\dots\alpha_n} = \frac{1}{\hbar^n}\frac{\partial^n H}{\partial k_{\alpha_1}\dots\partial k_{\alpha_n}}$ are the generalised $n$-th order velocity operators, $i\omega_\alpha$, $i\omega_\beta$ are Matsubara frequencies with $i\omega_\Sigma=i\omega_\alpha+i\omega_\beta$ and $[(\alpha,i\omega_\alpha) \leftrightarrow (\beta,i\omega_\beta)]$ indicates a symmetrisation between the two indices.
The shift current is considered in the case where $\alpha=\beta$ and $i\omega_\alpha = -i\omega_\beta = i\omega$.
Then, we notice that the first and third terms in the square brackets are off-resonant, thus, the contributions to the shift current are given by the second and fourth terms.
Focussing on the integrand of the final term with its symmetrised counterpart $\overline{\sigma}$,
\begin{equation}
    \overline{\sigma}=\sum_{\substack{a,b,c\\ c\neq a}}\frac{v^\mu_{ac}v^\alpha_{cb}v^\alpha_{ba}}{\varepsilon_{ac}}\left(\frac{f_{ab}}{\hbar\omega - \varepsilon_{ba} + i\eta}-\frac{f_{bc}}{-\hbar\omega - \varepsilon_{cb} + i\eta}+\frac{f_{ab}}{-\hbar\omega - \varepsilon_{ba} + i\eta}-\frac{f_{bc}}{\hbar\omega - \varepsilon_{cb} + i\eta}\right),
\end{equation}    
where we have analytically continued $i\omega\to\hbar\omega+i\eta$, $\eta$ being some infinitesimal quantity.
By appropriate exchange of indices, we obtain
\begin{align}
    \overline{\sigma} = &\hspace{3.4mm}\sum_{\substack{a,b,c\\ c\neq a}}\frac{v^\mu_{ac}v^\alpha_{cb}v^\alpha_{ba}}{\varepsilon_{ac}}\frac{f_{ab}}{\hbar\omega - \varepsilon_{ba} + i\eta}+\sum_{\substack{a,b,c\\ c\neq a}}\frac{v^\mu_{ca}v^\alpha_{ab}v^\alpha_{bc}}{\varepsilon_{ac}}\frac{f_{ab}}{\hbar\omega - \varepsilon_{ba} - i\eta} \nonumber
    \\
    &+\sum_{\substack{a,b,c\\ c\neq b}}\frac{v^\mu_{bc}v^\alpha_{ca}v^\alpha_{ab}}{\varepsilon_{bc}}\frac{f_{ab}}{\hbar\omega - \varepsilon_{ba} - i\eta}+\sum_{\substack{a,b,c\\ c\neq b}}\frac{v^\mu_{cb}v^\alpha_{ba}v^\alpha_{ac}}{\varepsilon_{bc}}\frac{f_{ab}}{\hbar\omega - \varepsilon_{ba} + i\eta},
    \label{eq:indexexchange}
\end{align}
where $a\leftrightarrow c$, $a\leftrightarrow b$, and $a\to c\to b\to a$ are performed in the second, third, and fourth terms respectively. We can now use
\begin{equation}
    \frac{1}{x\pm i\eta} = \frac{\mathcal{P}}{x} \mp i\pi\delta(x),
    \label{eq:sp}
\end{equation}
on \eqref{eq:indexexchange} and extract the resonant part to find
\begin{align}
    \overline{\sigma} &= i\pi\sum_{a,b}f_{ab}\left[\sum_{c\neq a}\frac{-v^\mu_{ac}v^\alpha_{cb}v^\alpha_{ba}+v^\mu_{ca}v^\alpha_{ab}v^\alpha_{bc}}{\varepsilon_{ac}}+\sum_{c\neq b}\frac{v^\mu_{bc}v^\alpha_{ca}v^\alpha_{ab}-v^\mu_{cb}v^\alpha_{ba}v^\alpha_{ac}}{\varepsilon_{bc}}\right]\delta(\hbar\omega-\varepsilon_{ba}).
    \label{eq:sigmabar}
\end{align}
By performing similar operations on the second term in Eq.~\eqref{eq:parkermorimoto}, with the aid of the Hellmann-Feynman theorem
\begin{equation}
    A^\alpha_{ab} = \frac{i\hbar}{\varepsilon_{ba}}v^\alpha_{ab},
    \label{eq:hf}
\end{equation}
one can reach the form of the shift current given in Eq.~\eqref{eq:scoriginal}.

We now move our attention to Eq.~\eqref{eq:scnumerical}.
By inspecting Eq.~\eqref{eq:parkermorimoto}, we notice that only $\overline{\sigma}$ contributes in the present situation, where our Hamiltonian is linear in $\vec{k}$, giving us
\begin{align}
    \sigma^\mu_{\alpha\alpha}(0;\omega,-\omega) &= \frac{i\pi q^3}{\omega^2}\int\frac{d^2\vec{k}}{(2\pi)^2} \sum_{a,b}f_{ab}\left[\sum_{c\neq a}\frac{-v^\mu_{ac}v^\alpha_{cb}v^\alpha_{ba}+v^\mu_{ca}v^\alpha_{ab}v^\alpha_{bc}}{\varepsilon_{ac}}+\sum_{c\neq b}\frac{v^\mu_{bc}v^\alpha_{ca}v^\alpha_{ab}-v^\mu_{cb}v^\alpha_{ba}v^\alpha_{ac}}{\varepsilon_{bc}}\right]\delta(\hbar\omega-\varepsilon_{ba}).
\end{align}
If we focus on the first term in the square brackets, we find that
\begin{align}
    -v^\mu_{ac}v^\alpha_{cb}v^\alpha_{ba} + v^\mu_{ca}v^\alpha_{ab}v^\alpha_{bc} &= -v^\mu_{ac}v^\alpha_{cb}v^\alpha_{ba} + \left(v^\mu_{ac}v^\alpha_{cb}v^\alpha_{ba}\right)^* \nonumber
    \\
    &= -2i\Im\left(v^\mu_{ac}v^\alpha_{cb}v^\alpha_{ba}\right).
\end{align}
We can perform the same operation on the second term which brings us to
\begin{align}
    \sigma^\mu_{\alpha\alpha}(0;\omega,-\omega) &= \frac{2\pi q^3}{\omega^2}\int\frac{d^2\vec{k}}{(2\pi)^2} \sum_{a,b}f_{ab}\Im\left[\sum_{c\neq a}\frac{v^\mu_{ac}v^\alpha_{cb}v^\alpha_{ba}}{\varepsilon_{ac}} + \sum_{c\neq b}\frac{v^\mu_{cb}v^\alpha_{ba}v^\alpha_{ac}}{\varepsilon_{bc}}\right]\delta(\hbar\omega-\varepsilon_{ba}),
\end{align}
where Eq.~\eqref{eq:scnumerical} can be reached by substituting for $q=-e$ and $\omega=\varepsilon_{ba}/\hbar$.
\end{widetext}

\section{Shift current at ($\theta$, $\Delta$) = ($2.0^\circ, 20~\rm{meV})$}\label{appx:extracalc}
\begin{figure}[t]
    \centering
    \includegraphics[width=8.5cm]{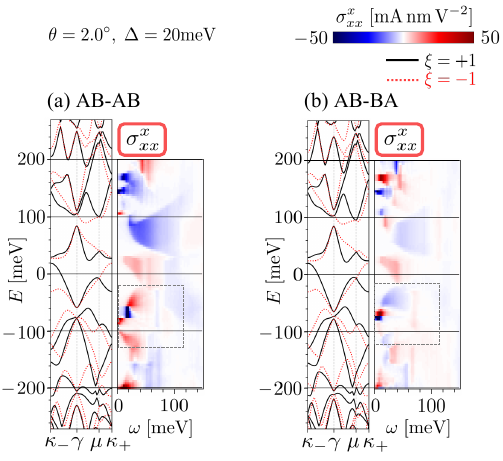}
    \caption{The band structures and $\sigma^x_{xx}(\omega;E_{\rm{F}})$ density plots of (a) AB-AB and (b) AB-BA stacked TDBG. Here, we used ($\theta$, $\Delta)$=($2.0^\circ$, 20~meV). In the region indicated by the grey dashed box, we find that the sign of the signal is reversed between the two variants.}
    \label{fig:A-tdbg20}
\end{figure}

In this appendix, we show the AB-AB and AB-BA TDBG band structures and the $\sigma^x_{xx}(\omega;E_{\rm{F}})$ density plots at ($\theta$, $\Delta$) = ($2.0^\circ$, 20meV) in Fig.~\ref{fig:A-tdbg20} (a) and (b), respectively.
The band structures for both TDBGs are slightly modified through the application of $\Delta$, however, the signal density plot in AB-AB TDBG remains relatively unchanged compared to the $\Delta=0$ case shown in Fig.~\ref{fig:3-tdbg04-40} (a).
For the AB-BA variant, where the signal vanished at $\Delta=0$, we now see a finite signal from the $C_{2y}$ symmetry-breaking provided by the vertical bias voltage.

Focussing on the region indicated by the grey dashed box, we can see that the sign of signal is reversed between the two TDBG stacking configurations.
This result provides additional support that the relative sign of the signal is inherited from the uncoupled TDBG, and that it can be seen in a range of twist angles.
We further comment that, as we increase the twist angle, the required value of $\Delta$ for the correlation to become visible becomes smaller, thus, easier to observe.
This can be understood in the following manner:
For the sign reversal to be observed, the low frequency contributions to the shift current must only come from transitions between neighbouring bands, which can be achieved when the bands are well separated.
Otherwise, when the bands are clustered, the resulting signal will be a sum of the contributions from several different transitions.
Then, cancellations between the contributions will ruin the correlation between the two variants.
In the case of $\theta=0.8^\circ$, a large $\Delta$ was required to separate the moir\'{e} subbands and to observe the negative correlation of the TDBG signals.
However, as the twist angle is increased, the bands gain more dispersion and become well separated, owing from the weakening of the moir\'{e} coupling.
Therefore, unlike the case for $\theta=0.8^\circ$, the role of $\Delta$ is solely to provide the symmetry breaking for a finite $\sigma^x_{xx}$ in AB-BA TDBG, and thus, a small $\Delta$ is enough to observe the sign reversal.

\section{Transformation of the conductivity tensor in uncoupled TDBG}\label{appx:transformation}

Here, we show the detail of the calculation that was performed to obtain Eq.~\eqref{eq:sigmatransformed}.
Using the explicit forms for the rotation matrix
\begin{equation}
    R(\psi) = 
    \begin{pmatrix}
        \cos(\psi) & \sin(\psi)
        \\
        -\sin(\psi) & \cos(\psi)
    \end{pmatrix},
\end{equation}
and the conductivity tensor in the coordinates aligned with the BLG lattice
\begin{equation}
    \sigma^\prime = 
    \begin{pmatrix}
        \sigma^{x^\prime}_{x^\prime x^\prime} & -\sigma^{y^\prime}_{y^\prime y^\prime} & -\sigma^{y^\prime}_{y^\prime y^\prime} & -\sigma^{x^\prime}_{x^\prime x^\prime}
        \\
        -\sigma^{y^\prime}_{y^\prime y^\prime} & -\sigma^{x^\prime}_{x^\prime x^\prime} & -\sigma^{x^\prime}_{x^\prime x^\prime} & \sigma^{y^\prime}_{y^\prime y^\prime}
    \end{pmatrix}
    , \label{eq:conductivitytensor}
\end{equation}
we can perform the matrix multiplication in Eq.~\eqref{eq:sigmatransformed} to obtain the $\sigma$ tensor in the unprimed coordinates.
It reads
\begin{equation}
    \sigma = R^{-1}\;\sigma^\prime\,(R\otimes R),
\end{equation}
where the $4\times4$ matrix $R(\psi)\otimes R(\psi)$ is explicitly given as
\begin{widetext}
    \begin{align}
        R(\psi)\otimes R(\psi) &=
        \begin{pmatrix}
            \cos(\psi) & \sin(\psi)
            \\
            -\sin(\psi) & \cos(\psi)
        \end{pmatrix}
        \otimes R(\psi) \nonumber
        \\
        &=
        \begin{pmatrix}
            \cos(\psi)\,R(\psi) & \sin(\psi)\,R(\psi)
            \\
            -\sin(\psi)\,R(\psi) & \cos(\psi)\,R(\psi)
        \end{pmatrix}
        \nonumber
        \\
        &=
        \begin{pmatrix}
            \cos^2(\psi) & \cos(\psi)\sin(\psi) & \cos(\psi)\sin(\psi) & \sin^2(\psi)
            \\
            -\cos(\psi)\sin(\psi) & \cos^2(\psi) & -\sin^2(\psi) & \cos(\psi)\sin(\psi)
            \\
            -\cos(\psi)\sin(\psi) & -\sin^2(\psi) & \cos^2(\psi) & \cos(\psi)\sin(\psi)
            \\
            \sin^2(\psi) & -\cos(\psi)\sin(\psi) & -\cos(\psi)\sin(\psi) & \cos^2(\psi)
        \end{pmatrix}
        .
    \end{align}
\end{widetext}
Then, the $\sigma^x_{xx}$ and $\sigma^y_{yy}$ components are given as
\begin{align}
    \begin{split}
        \sigma^x_{xx}(\psi) &= \sigma^{x^\prime}_{x^\prime x^\prime}\cos(3\psi)+\sigma^{y^\prime}_{y^\prime y^\prime}\sin(3\psi),
        \\
        \sigma^y_{yy}(\psi) &= -\sigma^{x^\prime}_{x^\prime x^\prime}\sin(3\psi)+\sigma^{y^\prime}_{y^\prime y^\prime}\cos(3\psi).
    \end{split}
\end{align}
Recalling that $\sigma^{x^\prime}_{x^\prime x^\prime}=0$ from the $M_x$ mirror symmetry, we finally end with
\begin{align}
    \begin{split}
        \sigma^x_{xx}(\psi) &= \sigma^{y^\prime}_{y^\prime y^\prime}\sin(3\psi),
        \\
        \sigma^y_{yy}(\psi) &= \sigma^{y^\prime}_{y^\prime y^\prime}\cos(3\psi),
    \end{split}
\end{align}
which is the expression we have used in Section~\ref{sec:blg}.

\bibliography{reference}
\end{document}